# Paraffin coated rubidium cell with an internal atomic vapor source

S. N. Atutov[*], A. I. Plekhanov, V. A. Sorokin

1 Institute of Automation and Electrometry SB RAS,
Koptyug Ave. 1, 630090 Novosibirsk, Russia
* atutov@fe.infn.it

S. N. Bagayev, M.N.Skvortsov, A.V. Taichenachev

2. Institute of Laser Physics SB RAS
Ac. Lavrentieva Ave.15B, 630090 Novosibirsk, Russia

**Abstract.** We present the results of a study on relaxation and diffusion processes of rubidium atoms in a rubidium cell with an internal vapor source. The cell is an evacuated glass bulb, which is characterized in that the source of atomic vapors in the form of a metal film Rb is evenly distributed throughout the inner surface of the bulb, and the paraffin film is uniformly distributed over the entire area and over the metal surface. By using laser optical pumping, we performed measurements of the relaxation time and the average number of bounces of optical pumped rubidium atoms in the bulb. We have measured the adsorption time of rubidium atoms by paraffin coating and rubidium atoms diffusion coefficient in paraffin used. A simple model of the pumping and atomic diffusion processes in the cell is discussed as well.

**1 Introduction**

Alkali atoms resonant cells are often used in compact atomic frequency standards. These devices exploit a microwave transitions through the ground states of the atoms to provide stable atomic frequency reference. At the base of the usual standard of frequency is glass cell filled with buffer gases. The buffer gases help prevent the penetration of polarized alkali atoms into the cell walls. It increases life - time of ground states atomic coherences in the cell which are the key to nowadays ultra-precise atomic clocks. Alternative and very effective approach to preserve atomic coherences is to use glass cells whose inner surfaces are coated with polymer anti-relaxation coating. This solution has several advantages, including extremely long term frequency stability over 40 years, a less sensitivity to magnetic field gradients and significantly narrow contrast resonance [1,2]. This kinds of coating was suggested for the first time in reference [3] and studied in references [4,5]. Currently, many publications deal with studies of different sorts of non-stick coatings (see Ref. [6] and references therein). Given the importance of the application of polymer coatings, extensive studies have been made of the polymers used, obtaining a large number of important and interesting results [7–11].

Conventional coated cell consists of the evacuated working volume that connected to the capillary containing at the end a small piece of alkaline metal which hold as the source for the saturated atomic vapor filling the cell. The number of polarized atoms in the working volume at the equilibrium is proportional to the life time of polarized atoms in the cell. It is known, that there are three main mechanisms that limit the life time of polarized atoms and thus population of polarized atoms in the cell. First one is depolarization of the atoms during of their collisions to the surface of the coating that is characterized by the time of preserving of the atomic coherences in the cell. Second limiting factor is leaking or escaping of polarized atoms from the working volume through the capillary towards to the piece of alkaline metal, where they finally relax (it is so called "reservoir effect", see, for example, Ref. [7, 12]). This mechanism is characterized by the average escape time it takes to lose polarized atoms through the capillary into the atomic source. The third mechanism is an adsorption of the polarized atoms by the coating that can be characterized by atomic storage time.

In the case of a cell, constructed as a spherical bulb with cylindrical capillary, the time of preserving of the atomic coherences is linear function of diameter of the bulb, while the leaking time of the atoms scales as a cubic of diameter of the bulb [13]. Under these conditions the escape time is relatively insignificant in big cells, but it becomes much more important in small cells, when the diameter of the bulb is the same order of diameter of the capillary. In small cell, escape of polarized atoms is dominating process and thus a number of polarized atoms is very small despite of a highest quality of the coating used. This puts a serious obstacle to create a super miniature atomic clock based on compact coated cell of conventional configuration.

In this article, we describe the first study on a rubidium cell with an internal vapor source. The cell is an evacuated glass bulb, which is characterized in that the source of atomic vapors in the form of a metal film Rb is evenly distributed throughout the inner surface of the bulb, and the paraffin film is uniformly distributed over the entire area and over the metal surface. We present the results of a study on the variation of the density of optically pumped atoms in the cell. These experimental studies are preceded by the discussion of a model of the pumping and atomic diffusion processes and by the definition of the relevant quantities.

## 2 Loading of a vapor cell

Let us consider a spherical cell with radius $R$ and with metal film of Rb that evenly distributed throughout the inner surface of the bulb, and the paraffin film is uniformly distributed over the entire area and over the metal surface. In absence of the reservoir effect and any other losses in the bulb, the density of rubidium $n$ is equal to the saturated density for rubidium atoms at a given temperature. Under these conditions, the loss rate of optically pumped atoms depends on two processes only: relaxation of pumped atoms on the surface of paraffin film and adsorption by metallic film of the pumped atoms which diffuse through the paraffin coating to the cell walls. The density of the pumped atoms in the cell reaches an equilibrium value $n_{pumped}$ when the sum of both loss rates of these atoms becomes equal to the production rate of the pumped atoms in the cell $P$ by the pump laser light:

$$P = \phi_{relax} + \phi_{adsorb} \tag{1}$$

where $\phi_{relax} = \dfrac{\pi R^2 \bar{v} n_{pumped}}{\chi_{relax}}$ is the loss rate of the pumped atoms on paraffin coating surface due to their relaxation, $\bar{v} = \sqrt{8kT/\pi m}$ is the mean velocity at temperature $T$ and $m$ is the of Rb atom, $\chi_{relax}$ is the average number of bounces it takes pumped atoms to relax on the surface; $\phi_{adsorb} = \dfrac{4\pi R^2 n_{pumped} D}{l}$ is the loss rate of the pumped atoms due to their adsorption by metallic film, $l$ is the thickness of the paraffin coating, $n_{pumped}/l$ - atomic gradient in the coating, $D$ - diffusion coefficient of the pumped atoms in paraffin. In the steady-state regime, the number of the pumped atoms in the bulb $N_{pumped}$ can be written in the following compact form:

$$N_{pumped} = P\tau_{total}, \qquad \tau_{total} = \dfrac{\tau_{relax} \tau_{adsorb}}{\tau_{relax} + \tau_{adsorb}}, \tag{2}$$

where $\tau_{relax} = \chi_{relax} l_{meanpass}/\bar{v}$ is the time of pumped atoms before being lost due to relaxation on the paraffin coating, $l_{meanpass} = 4R/3$ is the mean path of pumped atoms in a bulb,

$\tau_{adsorb} = lR/3D$ is the average time it takes of pumped atoms to be adsorbed by metal film deposited on the glass substrate of the cell walls.

As can be seen from Eq. 2 that the steady-state density of the pumped atoms is proportional to the life - time of the atoms in the bulb. The life-time and steady-state density are equal to zero at zero thickness of the paraffin film. The density increased linearly with increasing of the film thickness and, finally, it saturated at thick paraffin coating.

## 3 Experimental

We have manufactured several identical rubidium cells with different thickness of the paraffin coating. We use paraffin which consists of short chains of wide-range molecular weights. It has melting temperature ranging from 65 –75 $^0$C. The bulb of the cells (radius $R = 3$ cm) has a pump tube (A) and two sidearms (B,C) as it is shown in Figure 1. Each cell was connected to a turbo pump through a moveable vacuum stage which allows one to fix a cell in any position during films preparation.

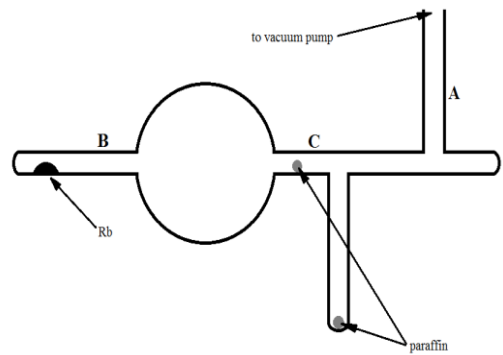

Figure 1.

The technique for preparing the paraffin-coated cells is discussed in [14]. Since our technique is somewhat different from the traditional one, therefore it is described here in detail.

Prior coating the bulb, sidearms and pump tube had been thoroughly cleaned of all traces of water and oxygen by continuous pumping for several days and by using an RF discharge of Ne. We consider the inner surface of the cell to have been cleaned when the discharge luminescence starts to assume a bright neon color and this color does not change remarkably after several hours of discharging. After that, cleaned cell was filled by neon at pressure slightly higher atmospheric pressure, sidebands were opened and a lump of natural-abundance Rb was inserted to the left sidearm, two pieces of paraffin were inserted into the right sidearm, as it is shown in Figure 1. Then both sidearms were closed, the cell was pumped out. The bulb with the exception both sidearms and pump tube was cleaned again by discharge of Ne.

To produce rubidium metal film, the left sidearm was heated up by flame to allow rubidium vapor to fill the bulb and condense on the wall. After that, the left sidearm containing a rest of rubidium was sealed off from the bulb at point B. Next, we have recorded the equilibrium rubidium pressure in the bulb for a given temperature that was done during several days. To make uniform paraffin coating across the bulb we have used the following procedure. We placed the cell in the vertical position with right sidearm up and allowed first piece of paraffin to fall into a rest of the left sidearm. Then we heated this piece to fill the bulb by paraffin vapor and allow the vapor to condense on the cell walls above the rubidium film. After that, we have recorded again the equilibrium rubidium pressure in the bulb. It was done during four days. Next, the cell was sealed off from the pump at point A. After that we allow the second piece of paraffin to fall inside the bulb and sealed off the right sidearm at point C. Then, we set the cell upside down in the vertical position and allow the second piece to fall back inside of a rest of the right sidearm. Finally we heated up second paraffin piece to fill the bulb by paraffin vapor again. Some times during paraffin deposition we cooled the bulb walls by liquid nitrogen and but obtained similar results.

Described procedure produces rather uniform paraffin coating without leaving any internal part of the bulb with naked metallic rubidium film or glass substrate. To make a paraffin film of desirable thickness in different cells we used paraffin pieces with calibrated sizes or we heated both pieces of paraffin with a calibrated time. The thickness of the paraffin film was estimated by a measured volume of distilled paraffin divided over the area of the whole coated surface. Note that a typical coating process for a one cell takes about ten days.

The setup for measurements of parameters of the cell is shown in Figure 2. The densities of the Rb vapor and the dynamics of the optical pumping process in the cell (1) are measured by fluorescence excited by a probe free--running laser (2). In certain optical pumping experiments, a separate pump free-running laser (3) is used. It has an expanded beam diameter of 10 mm. Fluorescence intensity is measured by a photodiode (FD) that is connected to a data acquisition system (DAQ).

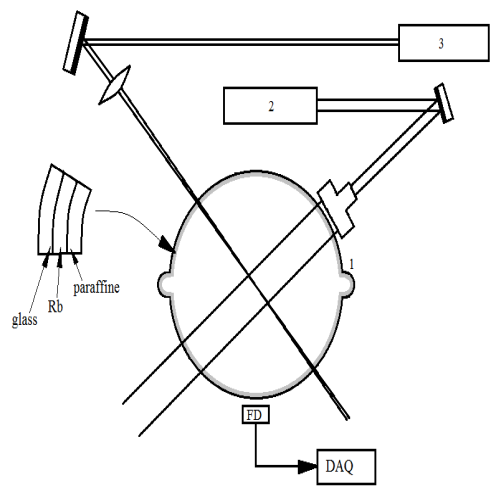

Figure 2. 1 – bulb, 2, 3 – diode lasers, FD – photodiode,

Because we used a non-stabilized probe laser, we caused the laser frequency to sweep periodically across the Rb optical transitions by a high frequency modulation (300 Hz) of the laser current. The density was extracted from averaged amplitudes of the fluorescence of all spectral lines of both isotopes of Rb atoms. To eliminate the influence of the probe laser on the dynamics of optical pumping, the laser beam was tightly focused inside of the cell. Due to beam focusing and frequency sweeping, Rb atoms interact with probe laser radiation a short time and this time is not enough to effectively pump the atoms. The described method allowed us to avoid any influence exerted frequency instability of the probe laser and optical pumping on the measurements of atomic density and of the life-time of the pumped atoms in the cell.

Figure 3 shows how the density of rubidium vapor depends on the duration of the coating process for one of the cells. It can be seen that prior paraffin coating, the saturated density of rubidium vapor in the bulb above of the naked rubidium film is constant with a small variation that is attributed to the variation of the ambient temperature. At room temperature (20 °C) the saturated vapor densities is ~ $6 \times 10^9$ cm$^{-3}$ [15].

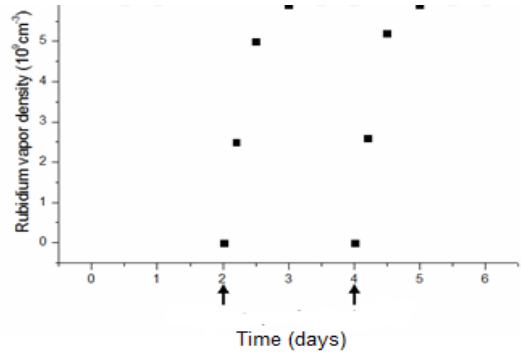

Figure 3. Rubidium vapor density versus of time.

The density sharply drops at the moment of time of starting of the paraffin deposition above the rubidium film that is indicated by the first arrow. After about 4 days, the density approaches a limit, which is exactly equal to the saturated rubidium vapor density. The density sharply drops again at the moment of starting of the second paraffin deposition that is indicated by another arrow. Again after several days the density approaches the same limit of the density. Note, that this curve is identical for all cells studied; only the time of rubidium density recovering is slightly shorter for a cell with a thinnest paraffin coating.

Initially, at both points indicated by the arrows, the density of the alkali metal vapor in the bulb is extremely small, so that fluorescent light is not observed. This can be explained by the fact that the new coating is chemically active due to gases such as oxygen or water dissolved in paraffin. Rubidium atoms slowly diffuse through the paraffin coating from the metal film towards the vacuum volume of the bulb. Diffuse atoms chemically react with impurities in the paraffin coating and, as a result of this process, the atomic density inside the bulb gradually increases. Density approaches the saturated vapor density when all impurities are neutralized and the coating becomes chemically inert and pure. Due to the absence of any losses in the bulb, the density of rubidium at limit is exactly equal to the saturated density for rubidium atoms at a given temperature. This contrasts to the case of coated cell of the conventional configuration, where the density of vapor atoms is 10-70% lower than density of saturated vapor of alkali metal, even in properly purified and cured cells [see, for example, 14]. Low saturated vapor density in conventional cells attributable to the irremovable loss of atoms via the adsorption of the atoms onto the glass substrate of the coating. The atoms from the source (where the atomic density is maximal) diffuse through the capillary to the cell vacuum volume. Then the atoms, after many bounces in the cell, start to diffuse deep inside the coating towards the glass substrate surface, where they are irreversibly absorbed. Not that, it takes rather long time to collect several equilibrium mono-layers of atoms on the glass substrate and obtain truly steady-state atomic density in common resonant cells [13, 16].

To characterize the quality of the manufactured cells, we performed measurements of the relaxation time of optical pumping $\tau_{relax}$ and absorption time $\tau$ of Rb atoms by the coating. We have used the results of measurements to evaluate the average number of bounces of the optically pumped atoms in the cells $\chi$ and diffusion coefficient of rubidium atoms in paraffin.

To make sure that there is no rubidium on surface of the paraffin film, we employed particles photo-desorption effect [14,17.18]  At time $t = 0$, we illuminated the bulb by a photographic flash lamp and recorded the fluorescence of Rb atoms. Figure 4 shows the fluorescence signal of the atoms in the bulb as a function of time.

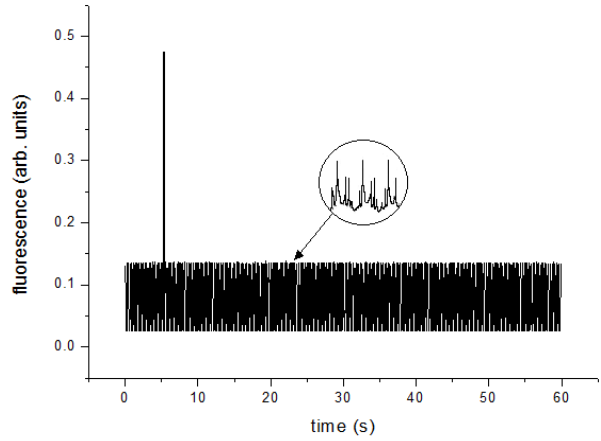

Figure 4. Fluorescence signal of the atoms
in the bulb as a function of time

It is seen that with the exception of the peak due to a flash of light hitting the photodiode, no increase in the density of rubidium caused by photodesorption was detected. This means that there is no rubidium either on the surface of the paraffin film or in the volume of the film close to its surface.

To characterize the quality of the paraffin used, we performed measurements of the relaxation time of the optical pumped Rb atoms $\tau_{relax}$ and the average number of bounces of the optically pumped atoms in the bulb $\chi_{relax}$. The laser frequency of the pump laser 2 was tuned to be resonant to 5S1/2 ($F = 3$) $\rightarrow$ 5P3/2 ($F' = 2, F' = 3, F' = 4$) of $^{85}$Rb optical transition, until a maximum of fluorescent intensity in the separate, uncoated cell was obtained. Then the pump beam in the bulb was abruptly opened by a shutter. The atoms on the Rb atom ground state 5S1/2 ($F = 3$) were excited by the radiation; and the radiation populated 5S1/2 ($F = 2$) ground state through the intermediate atomic upper states 5P3/2 ($F' = 1\ F' = 2, F' = 3, F' = 4$). As a result of the optical pumping process, the amplitude of 5S1/2 ($F = 3$) line decreased while the amplitude of 5S1/2 ($F = 2$) line increased (see Figure 5).

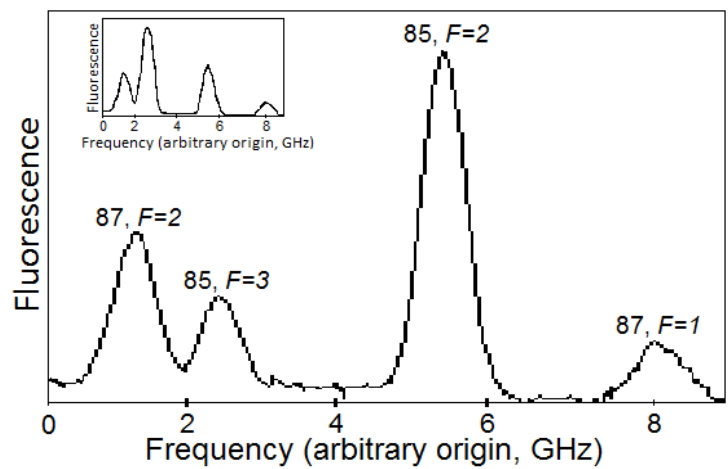

Figure 5. Spectrum of the pumped rubidium atoms.
The inset shows not pumped rubidium spectrum.

Then the pump beam was blocked and the population on the 5S1/2 (F = 3) level as a function of time was recorded (see Figure 6). The curve on Figure 6 is described by two exponentials, with fast and slow times of 4 and 1,8 seconds, respectively. Taking the maximum lifetime 1,8 second, bulb radius $R = 3$ cm, the mean path inside the bulb $4R/3 = 4$ cm, Rb thermal velocity $\bar{v} = 2.7 \times 10^4$ cm/s, we found the number of bounces $\chi_{relax} = 1,2 \times 10^4$. Note, the number of bounces measured in this experiment is typical for our paraffin used.

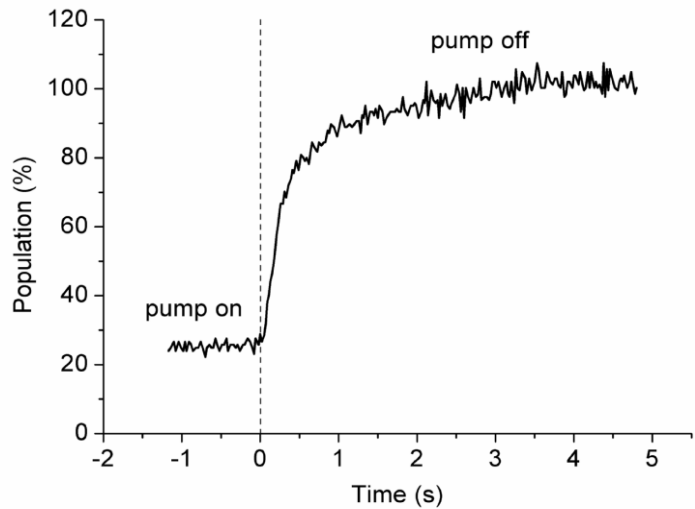

Figure 6. Population on the 5S1/2 (F = 3) level as a function of time.

We attempted to measure the life time of the pumped atoms in the cell as a function of paraffin coating thickness. The experiment was made in one of the cells. Starting with a bare metal film we heated the pieces of paraffin in both sidearms with a calibrated time. This allowed us to get a paraffin film of different thickness. The results of measurements are shown on Figure 7. Note that X axis in the ln scale.

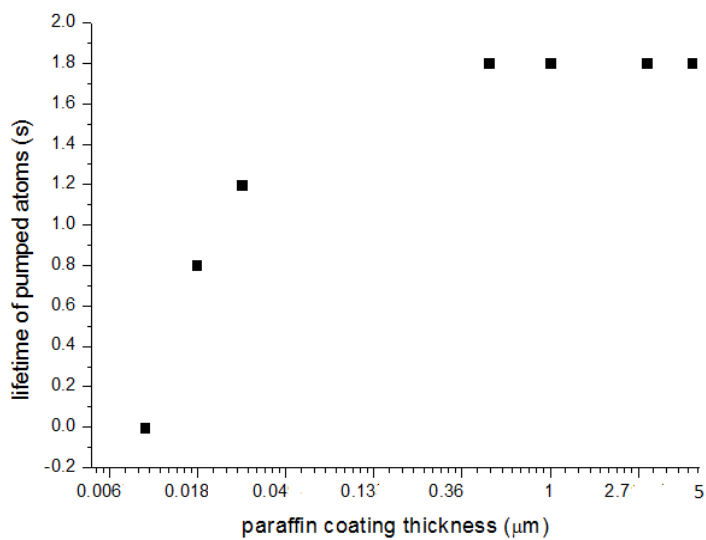

Figure 7. The life time of the pumped atoms in the cell as a function of paraffin coating thickness

It can be seen that the life time of the pumped atoms in the bulb depends to the thickness of paraffin coating: without coating the relaxation life - time is close to zero, after which it increases linearly as the thickness increases and it tends to saturate at a thickness of about of 0.1 μm. The form of the graph agrees with the model (Eq. 2).

Then we measured the absorption time $\tau_{adsorb}$ of Rb atoms by the coating. At time $t = 0$, we locally heated up a small part of the bulb wall by a hot glass bar during one second. The warm and narrow point on the wall created a burst of rubidium atoms that penetrate through the paraffin coating to the bulb volume. Then, we recorded the fluorescence decay that was due to adsorption of Rb atoms back by the coating. Figure 8 shows the typical dependence of the fluorescence intensity of atoms to time. One can see that the intensity of the fluorescence of the Rb atoms during the heating (a peak at $t = 0$) increases sharply and then, after approaching a maximum, the intensity slowly declines. Using the decay of the fluorescence intensity we measured the absorption time of Rb atoms $\tau_{adsorb}$. The time was found to be equal 600 seconds for the paraffin coating thickness of 5 micrometers.

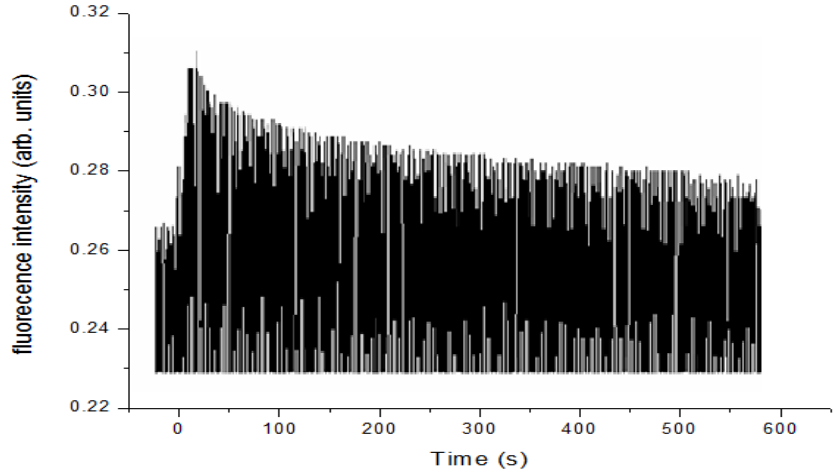

Figure 8. Fluorescence intensity of atoms in the bulb
as a function of time (paraffin coating thickness 5 micrometers)

Figure 9 shows the dependence of the absorption time $\tau_{adsorb}$ on the paraffin film thickness $l$.

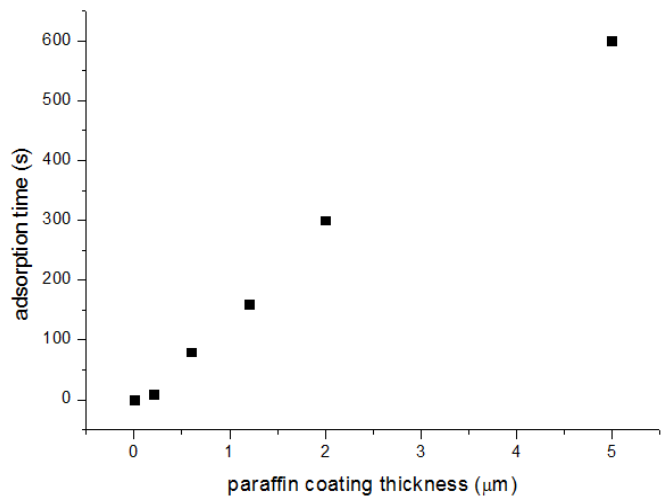

Figure 9. Dependence of the absorption time $\tau_{adsorb}$
to the paraffin film thickness $l$.

We use the plot presented in Figure 9 to determine the diffusion coefficient of the Rb atom in the paraffin film. This is achieved by fitting of relation $D = lR/3\tau_{adsorb}$ to the experimental points. The average value of the diffusion coefficient of rubidium atoms $D$ for $R = 3$ cm was found to be equal to $8 \times 10^{-7}$ cm$^2$/s. The statistical uncertainty is ±30%, mainly attributable to an uncertainty in the measurements of the thicknesses of the coating. The value of the diffusion coefficient measured is equal within an error bar to the paraffin diffusion coefficient $5 \times 10^{-7}$ cm$^2$/s that was measured in Ref. [13] and to the paraffin (*n*-Octadecane) self-diffusion coefficient $4.6 \times 10^{-7}$ cm$^2$/s that was measured in reference [19].

Conclusion

An experimental study of relaxation and diffusion of optically pumped Rb atoms in the cell with internal vapor source was presented. We verified the absence of any losses of atoms in the bulb and that the density of the atoms in the vapor is exactly equal to the saturated density for rubidium atoms at a given temperature. The absence of chemical reaction of Rb atoms with coating and is supported by the fact that, for example, the time of relaxation of the pumped atoms does not depends to the time of employment of the cell. We have measured the relaxation time of the optical pumped Rb atoms and the average number of bounces of the optically pumped atoms in the bulb, and show that the number of bounces measured in this experiment is typical for our paraffin used. By comparing cells manufactured one year ago we verified the long term stability of our cells with internal atomic source.

These results are the basis for the development of miniature spherical cells. Recently, this kind of miniature 5 mm in diameter cells was successfully used for development of compact atomic clocks at Institute of Laser Physics SB RAS.

We would like to thank Prof. A. Shalagin, Prof. N. Surovtsev for stimulating discussions. Special thanks for D. Shadrin for the help in experiments and R. Robson (McKillop) for careful reading of the manuscript.